\documentclass{aimc2026}

\newif\ifshellesc
\shellesctrue  %

\newif\ifanonymize
\anonymizefalse   %

\usepackage[utf8]{inputenc} %
\usepackage[T1]{fontenc}    %
\usepackage{hyperref}       %
\hypersetup{
  colorlinks = true,  %
  urlcolor   = blue,  %
  linkcolor  = black, %
  citecolor  = black  %
}
\usepackage{url}            %
\usepackage{booktabs}       %
\usepackage{amsfonts}       %
\usepackage{amsmath}
\usepackage{nicefrac}       %
\usepackage{microtype}      %
\usepackage{graphicx}       %
\usepackage{cleveref}       %
\usepackage{layout}
\usepackage{calc}
\usepackage{xcolor}
\usepackage{flushend}
\usepackage{subcaption}

\usepackage{siunitx}
\title{Rescuing Performance from the Demo: Co-Designing Drum Gesture Mappings with a Percussionist}

\author{%
 Jordie Shier \thanks{\texttt{j.m.shier@qmul.ac.uk}} \\
 Centre for Digital Music\\
 Queen Mary University of London\\
 London, UK \\
 \And
 Teresa Pelinski \\
 Centre for Digital Music \\
 Queen Mary University of London \\
 London, UK \\
 \AND
 Charalampos Saitis \\
 Centre for Digital Music \\
 Queen Mary University of London \\
 London, UK \\
 \And
 Andrew Robertson \\
 Ableton AG \\
 Berlin, Germany \\
 \And
 Andrew McPherson \\
 Dyson School of Design Engineering \\
 Imperial College London \\
 London, UK \\
}

\begin{document}

\twocolumn[%
  \maketitle
  \begin{abstract}
Augmenting instruments with sensors and neural network mappings is a well-explored digital musical instrument design approach. While augmentations can create new expressive opportunities, they also exert aesthetic influence and can constrain musicians' gestural language, which, if left unchecked, can lead to technological capture. To examine this, we conducted a study with a professional percussionist, co-developing a gesture mapping toolkit and recording a ten-track album. Drawing on the concept of productive dissonance, our study aimed to hold the musician's aesthetic in tension with technological constraints. This, along with a practice-based reflective approach, supported the development of a continuous gesture recognition method for percussive mapping and surfaced insights into the design process. We identify knowing-when as a form of tacit knowledge that supported productive dissonance, and raise an open question: absent a musician's broader social context, how do we know whether a technology's influence is genuinely supporting their practice?
  \end{abstract}%
]

\aimcnotice %

\section{Introduction}
\label{sec:intro}

Designing digital musical instruments (DMIs) often involves creating mappings between sensors and audio synthesis algorithms~\citep{hunt_importance_2003}, allowing for gestural manipulation of diverse sound sources.
Machine learning and artificial intelligence are increasingly being used to create these mappings, supporting complex relationships between input  and sound generation~\citep{jourdan_machine_2023}.
When applied to existing instruments, such mappings can allow musicians to ``recycle'' their existing musical skills within a new timbral sound world~\citep{tremblay_surfing_2010}.

However, recent critical discourse has raised the risk of reification of mapping, where the metaphor of mapping comes to define musical performance itself~\citep{mcpherson_mapping_2024}.
In this paper, we examine this risk and identify a concrete consequence, which we call \textit{technological capture}, where the constraints of a mapping system overwhelmingly shape musical practice and performance, ultimately preventing musicians from ``recycling'' their musical skills.

To engage with this tension, we propose a bespoke co-design approach strongly influenced by technical practice research (TPR)~\citep{pelinski_ways_2025}.
More specifically, a researcher (the first author) worked closely with a professional percussionist over the course of four months to co-develop  a drum gesture mapping toolkit and incorporate it into the percussionist's practice.
Creating and recording a musical performance with our co-designed mappings was a key goal, providing a musical anchor to technical practices.

The contributions of this paper are twofold. First, as outcomes of the co-design, we share the co-developed drum gesture mapping toolkit built in Max4Live and a ten-track album presented as an annotated portfolio \citep{gaver_annotated_2012} on an accompanying website.\footnote{\url{https://jordieshier.com/projects/aimc2026/}}
The toolkit contributes a novel method for real-time mapping of continuous percussive gestures (e.g., brushing) to synthesizer parameters using neural networks.
Second, we share insights from attempting to put the critical discourse on mapping into practice: our collaboration was designed to work against the risk of technological capture by holding a professional musician's aesthetic in tension with the constraints of our system.
Through documentation and reflexive practices, we identify a form of judgement that supported sustaining this tension, which we characterise as \textit{knowing-when}~\citep{ryan_knowing_2009}: tacit knowledge of when to attend to the technology and when to the music.
Our study also surfaced a limitation of the laboratory setting itself\textemdash we reflect on this in our discussion and present it as an open question to the community: absent a musician's broader social context, how do we know whether a technology's influence is genuinely supporting their practice?

\section{``Rescuing the performance from the demo''}
\label{sec:design-background}

The provocation ``rescue performance from the demo'' offered by \citet{norman_nime_2025} highlights a key tension in designing and performing with new technology for music, including DMIs.
Is the performance about the music or about demonstrating the affordances of the new technology?
These goals are often at odds with each other in an unresolvable way.
We argue that this tension is actually a positive, echoing what \citet{mcpherson_mapping_2024} describe as ``productive dissonance'' and the notion of \textit{dissensus}~\citep{norman_nime_2025}. 
It is when this tension resolves in the direction of technology that the driving aesthetic force behind a musical performance is subordinated by the goals of technology development.
We borrow the term \textit{capture} from regulatory theory \citep{dal_bo_regulatory_2006}, where it describes the process by which a regulatory body comes to serve the interests of the industry it was designed to regulate. By analogy, \textit{technological capture} describes the process by which musical practice comes to serve the interests of technology development rather than the other way around.

Gestural control in musical interaction design offers a useful reflection on when and how ``technological capture'' manifests. 
\citet{tuuri_who_2017} describe the \textit{choreographing effects} of technology on musical gesture, querying whether it is the musician who controls the technical system or the other way around.
They propose two modes of technological choreographing: \textit{push and pull effects}.
An example  was demonstrated by~\cite{jack_rich_2017} in a performance study with a digital percussive instrument, 
where musicians discarded gestures that produced no sonic result on the instrument (push effects) and converged on the most ergonomically convenient gestures that achieved sonic results (pull effects).
Jack et al. use the term \textit{bottleneck} to characterise attributes of a DMI that act to constrain a gestural language and steer a musician to play in certain ways.

Bottlenecks in DMI mapping are largely unavoidable due to the often necessary process of representing a musical performance in a computationally efficient form.
Representations are useful for analysing phenomena~\citep{mcpherson_mapping_2024, saitis_constructing_2025} and as input to machine learning algorithms, but they can become problematic when a reduced symbolic representation of music is inverted to become its generative basis.
For example, if a DMI applies an onset detector to an audio stream, then it presupposes that ``onset'' is a musically meaningful concept, even though onsets might not be a stable, pre-existing entity in the absence of an onset detector~\citep{reed_shifting_2024}.
A musical worldview containing onsets can be aesthetically desirable in certain contexts: for instance, a kick drum trigger is often used in metal performance to achieve timbral consistency and clarity during rhythmically fast or complex sections~\citep{weatherhead_drum_2025}.
But to assume it is universal to all music would be problematic.
This underscores the importance of contextualising technical systems within a ``well-defined musical aesthetic''~\citep{mcpherson_mapping_2024}, where unsuitable representations can generate productive dissonances to guide design. 

To unpack these tensions, our study centres on directly collaborating and co-designing with a professional musician to provide a well-defined musical aesthetic to work against.

\section{Methodology}
\label{sec:design-method}

In this section, we summarise how the study-collaboration was structured around co-design and practice research, and provide details on documentation and reflection. 

\subsection{Study Co-Design}
\label{sec:study-design}

The study was conducted over four months at Queen Mary University of London qMedia Studios, structured in three phases.
The first phase consisted of a semi-structured background interview and initial co-design session, during which the researcher learned about the musician's broader practice and aesthetic background, introduced the technical starting point, and conducted an initial brainstorming session to identify musical directions.
The second phase consisted of five co-design sessions (2.5 hours each), spaced approximately two weeks apart, interleaved with periods of technical development by the researcher. 
Each session followed a consistent structure: a brief semi-structured reflection on the previous session, a co-design working block focused on developing and refining a mapping for a specific musical context, recording a short musical performance, and a reflection on the performance to set goals for the next sprint. 
The third phase consisted of two full recording days structured to replicate a real studio session, with making music, rather than technical development, as the explicit driving goal to anchor the co-design in a compositional aim.
Thoroughly documenting the study was an important consideration to support reflexive practices, and is detailed in Section~\ref{sec:method-documenting}.
The musician was paid at the local union hourly and daily rates throughout.

\subsection{The Participants-Collaborators}
The study was conducted by the first author (referred to as \emph{the researcher}) and Jem Doulton (referred to as \emph{the musician}), who was hired as a musical collaborator.
The study originated after the researcher was introduced to the musician at a concert; the musician expressed interest in the researcher's work on percussive gesture mapping, and the study design (detailed in Section~\ref{sec:study-design}) was planned collaboratively over subsequent conversations.
Background information on the musician (collected from the background interview) and researcher is provided here to situate their collaboration and individual aesthetic background.

\subsubsection{The Musician}
Jem Doulton is a professional percussionist working in London, UK, playing in multiple musical projects ranging from folk country to free improv.
He describes his aesthetic as occupying a space between rock and jazz, both sonically and in its energy and attitude, in his own words, music that has a ``fuck you'' quality.\footnote{Henceforth, all literal quotes are from the collected study documentation unless indicated otherwise.}
The musician considered tacit knowledge a key element of his practice, for instance in the selection of cymbals, drums, sticks, and other ``trinkets'' to achieve a desired sonic result during a performance.

Jem put special value on the dexterity and ``hands-on'' character of his practice, an aspect which became a touchstone in the co-design study, which he described as ``marrying the tech and the human.''
In his past experience, incorporating digital technology into his practice had sometimes worked against this dexterity. In particular, he described an experience of playing in a band where over the course of several years his full acoustic drum kit was slowly replaced by electronic drum pads to the point where he felt he could have been replaced by a backing track:
``I felt trapped ultimately, and that is where I think you can take the tech a little bit too far.''
During the co-design study we repeatedly returned to this anecdote as an example of what we wanted to avoid\textemdash the ``technological capture'' we have described in Section \ref{sec:design-background}.

\subsubsection{The Researcher}
The researcher, who is the first author of this paper, has a background in computer science and received formal training in audio engineering, and music theory/composition at an undergraduate level in a contemporary classical context.
Prior to academia, he was a professional musician and performed and produced electronic music in several collaborative projects.
In his primary musical project, Napoleon Skywalker, he produced and performed electronic dance music using Ableton Live and analogue synthesizers, and performed alongside a live drummer.
Sound design and synthesizer programming are central to his musical practice, influenced by experimental electronic dance music and hip-hop.
This practice has directly informed his academic research on real-time percussion-to-synthesizer timbre mapping and DMI design more generally.
His technical research in this area includes the application of deep learning and evolutionary algorithms to estimate synthesizer parameters and to facilitate real-time, audio-driven interaction.

\subsection{Documenting and Reflecting}
\label{sec:method-documenting}

In this study-collaboration we followed a practice research approach; this means that our principal research method is the collaborative design \emph{practice} itself \citep{Bulley2021a}. Practice research is not uncommon in instrument design \citep{Bowers2018, Johnston2016a, Pelinski2026} or generally in design \citep{bang_designing_2023,Gaver2012, Devendorf2020a} or the arts \citep{Borgdorff2010, Spatz2015}. \emph{Practice research} is however a broad term; here we draw primarily from the technical practice research (TPR) that  \citet{pelinski_ways_2025} propose, also inspired by reflective design \citep{Sengers2005} and feminist HCI \citep{rode_theoretical_2011}. TPR emphasises documenting and reflecting on the practice as closely as possible to the moment of it; seeking to capture the mutual mediations between technical development, aesthetics, and social dimensions; and, importantly, their evolution as the process unfolds.
\citeauthor{pelinski_ways_2025} also point out the relevance of documenting and reflecting on the technical development as it unfolds; as the more technical phases of instrument design are often disregarded in practice research approaches to DMI design.  

The practice was documented in a variety of forms. All studio sessions were audio and video recorded, including the interactions between researcher, musician, and studio visitors, which were transcribed and later thematically coded~\citep{braun_reflecting_2019}; this was also the case for all interviews conducted outside of the studio sessions. Technical development was documented through version control (git) and through the researcher's design journal. 

We followed several parallel strategies to foster reflection: The researcher's design journal served not only to document the technical process but also for the researcher to reflect on it as it unfolded. git commit messages were used in a similar way. Between sessions, the researcher went through the previous sessions' documentation, incorporating the musician's references and aesthetic dimensions into the technical development loops.
The researcher also encouraged the musician's reflection through questions asked at the beginning of each session and after periods of performance to query their felt experiences, inspired by \cite{petitmengin_describing_2006}.
Additionally, the second author conducted five semi-structured interviews with the researcher periodically throughout the study, asking questions related to the progress of technical development and prompting reflection about the evolution of the co-design itself. These interviews served as a form of dialogic reflection and also captured ``snapshots'' of the process through the viewpoint of the researcher. The insights presented in the later sections are the product of revisiting the collected documentation and further articulating these reflections.

\section{Toolkit and Performance}
\label{sec:toolkit}

Augmenting percussion instruments with sensors and mappings to synthesizers has been explored in both commercial tools such as Sensory Percussion\footnote{\url{https://sunhou.se/sensorypercussion}} and open-source systems such as DataKnot.\footnote{\url{https://github.com/rconstanzo/data-knot}} Both tools support percussive gesture recognition, including onset detection, classification, and mapping of drum strokes to samples or synthesis parameters.
The technical starting point for this study was a set of gesture mapping tools built on DataKnot and developed by the researcher~\citep{shier_real-time_2024, shier_designing_2025}.

\subsection{Gesture Mapping Toolkit}
\begin{figure*}[htpb]
    \centering
    \includegraphics[width=0.9\linewidth]{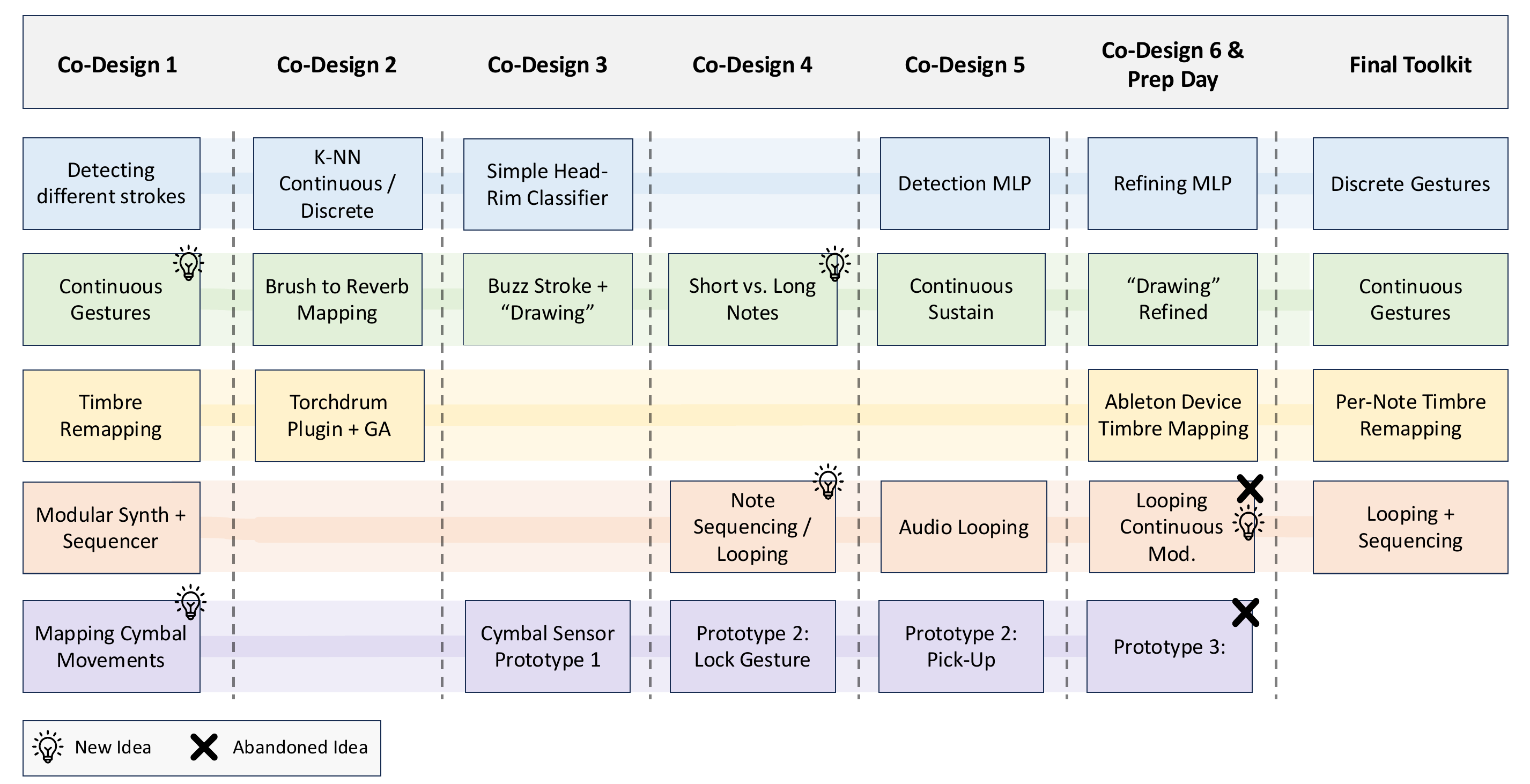}
    \caption{Key developments and insights throughout the various co-design sessions, and what elements of the toolkit ended up being included in the final kit. A cymbal sensor was explored, but didn't make it into the final kit. Section~\ref{sec:cymbal-sensor}  provides additional information on the cymbal sensor development and subsequent abandonment.}
    \label{fig:toolkit-timeline}
\end{figure*}

\begin{figure*}[htpb]
    \centering
    \includegraphics[width=0.9\linewidth]{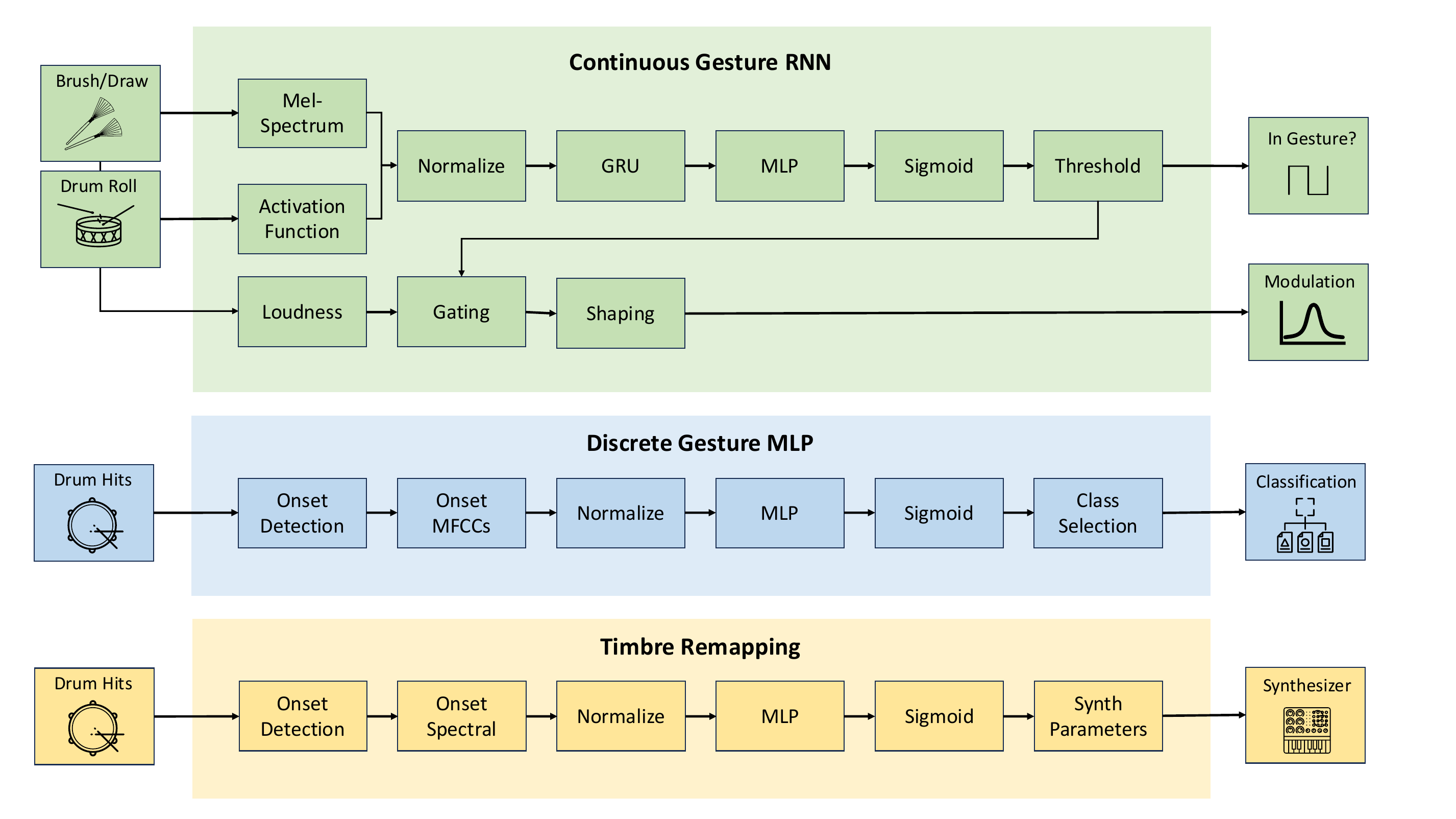}
    \caption{Overview of the three gesture recognition systems: the continuous RNN, discrete MLP, and timbre remapping. The continuous gesture RNN was trained on either brushing/drawing gestures or to detect buzz rolls---these are separate networks, but share the same architecture. The output of the continuous RNN is a binary signal indicating whether a continuous gesture has been detected and a continuous modulation signal. The discrete MLP is trained as a multiclass classifier based on different types of drum hits (i.e., head vs. rim). Timbre remapping uses a neural network trained as a regressor on pairs of spectral audio features and synthesizer parameters for a specific synthesizer. The recognition algorithms each support a different style of playing and were combined in various ways across the album recording.}
    \label{fig:neural-overview}
\end{figure*}

The toolkit can be broadly divided into two groups: objects that perform gesture recognition on streaming percussive audio and map the resulting features to synthesizer parameters, and objects that support musical structure within a live performance context such as audio looping and MIDI sequencing. Figure~\ref{fig:toolkit-timeline} provides a visual overview of the stages of development throughout the co-design timeline and Figure~\ref{fig:neural-overview} overviews the neural network mapping systems described in the following subsections.
Full technical details are provided in Appendix~\ref{app:tech-details}.

\subsubsection{Discrete Gesture Recognition}
The discrete gesture recognition component uses onset detection to determine when a drum has been struck, followed by audio feature extraction and classification with a lightweight MLP. This is the standard approach in existing percussive mapping tools---onset detection provides a reliable, low-latency event trigger, and classification between playing zones such as drum head and rim was sufficient for the performances we developed. However, as we discuss in Section~\ref{sec:tracing-continuous-gesture}, this approach carries a representational assumption: it renders all percussion as a stream of discrete events, which is not universal to all playing techniques or musical contexts.

\subsubsection{Continuous Gesture Recognition}
A significant portion of the study was devoted to extending the toolkit to support continuous gesture recognition, which involves mapping gestures that sustain over time rather than triggering discrete events. 
To our knowledge, recognising a sustained gesture as a category distinct from discrete strikes and using it to drive control-rate parameters of a parametric synthesizer has not been implemented in a real-time percussive performance context.\footnote{Sensory Percussion has a continuous interpolation mode; however, it is still based on a sequence of discrete onset triggers.}
We focused on two gesture types. The first were sustained gestures such as wire brush strokes on a drum membrane. An initial NMF-based approach \citep{lee_learning_1999, smaragdis_non-negative_2003} was explored to isolate brush-like spectral components, but was unable to reliably exclude discrete strikes. We subsequently trained a recurrent neural network (RNN) — specifically a gated recurrent unit (GRU) \citep{cho_properties_2014} following the architecture used by \citet{engel_ddsp_2020} — on manually labelled recordings from the co-design sessions, using a Mel-scaled spectral input representation. The second gesture type was buzz stroke rolls. Here the same RNN approach was applied using amplitude envelope features as input, trained to distinguish buzz strokes from regular double strokes.
Both models were implemented for real-time inference using the RTNeural plugin for Max/MSP,\footnote{\url{https://github.com/spluta/RTNeural_Plugin}} built on RTNeural~\citep{chowdhury_rtneural_2021}, with frame-based features mappable to parameters in Ableton Live.

\subsubsection{Timbre Remapping and Performance Devices}
The toolkit also includes objects for timbre remapping~\citep{stowell_timbre_2010, shier_designing_2025}, from percussive input to synthesizer parameters, and performance devices including audio looping and melodic MIDI sequencing. These are described in more detail in Appendix~\ref{app:timbre-remapping}.

\subsection{Musical Performance}
\begin{figure}[htpb]
    \centering
    \includegraphics[width=1.0\linewidth]{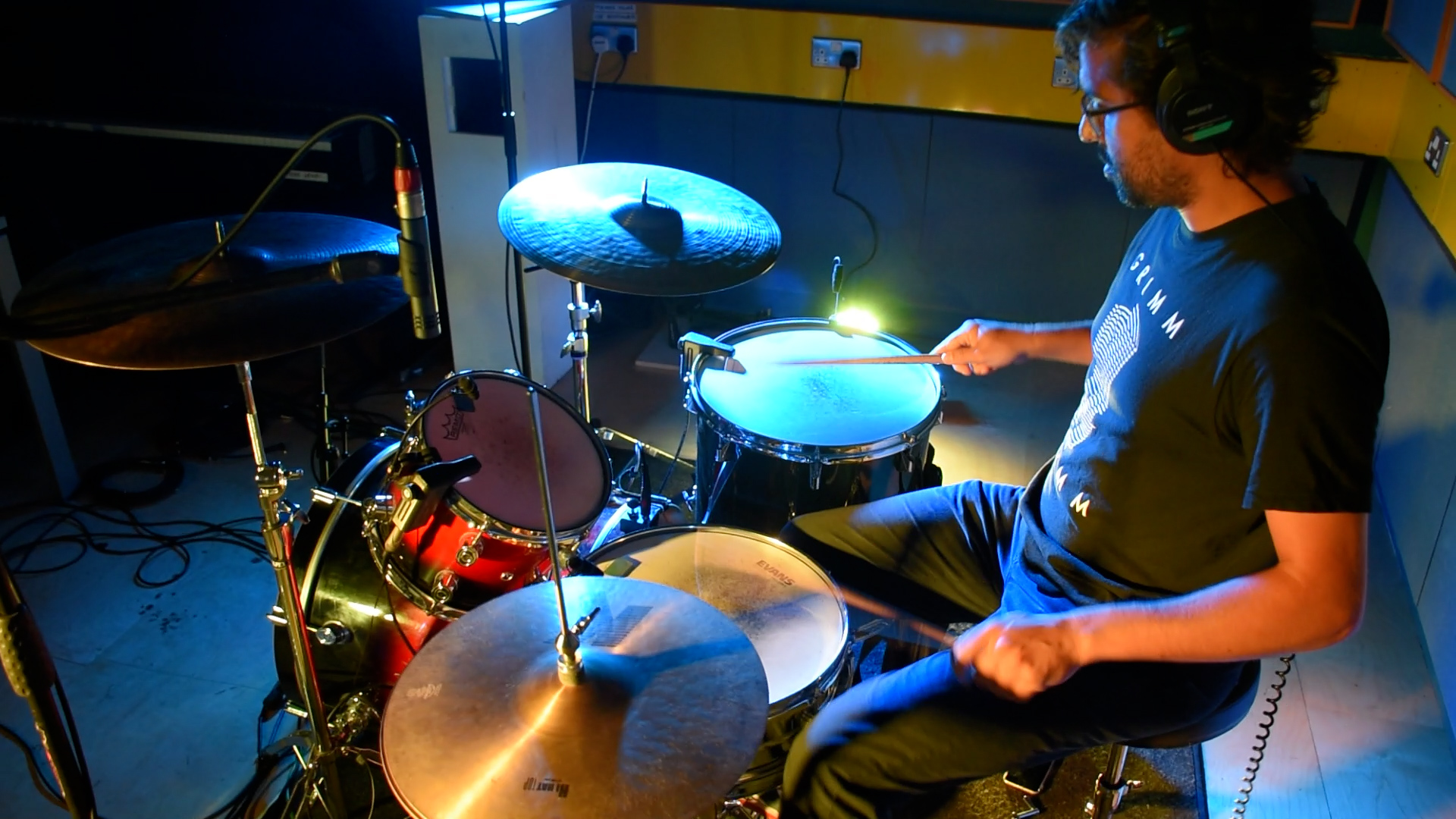}
    \caption{Performance photo of the musician playing the augmented drum kit designed as a part of this study.}
    \label{fig:musician}
\end{figure}

The aesthetic direction for the album emerged early in the collaboration. During the first session the musician mentioned a previous solo project, a drum and electronics album he had made as a soundtrack to a play,\footnote{Short Stories by Jem Doulton:~\url{https://jemdoulton.bandcamp.com/album/short-stories}} and noted that he had ``always wondered if I'd ever do another [one].'' This reference lay dormant until much later in the study, when the format of the recording crystallised: short, dense grooves\textemdash ``a beat tape''\textemdash with heavy synthesizer sounds that ``could potentially be danced to.'' The resulting album, recorded over two full days structured to replicate a real studio session, consists of ten tracks and constitutes the musical output of the study. It is presented as an annotated portfolio \citep{gaver_annotated_2012} with edited video performances available on the accompanying website. A public release with professional mix and mastering is planned.
Figure~\ref{fig:musician} shows a photo taken from the recording day of the musician performing.

\section{Productive Dissonance in Practice}
\label{sec:productive-dissonance}

In this section we present design arcs from the study-collaboration that illuminate where productive dissonance was sustained and where it surfaced limitations.

\subsection{Representational Dissonances}
\label{sec:rep-diss}

Representational dissonances arise when the computational representation of gesture imposed by a mapping system fails to capture the range of a musician's embodied practice. 
It describes the gap between the computational representation of a gesture (what the system can ``see'') and the gesture itself (what the musician is actually doing).
The following arcs trace how this gap surfaced through contact with practice, and how we designed in response to this dissonance.

\subsubsection{Continuous Gestures}
\label{sec:tracing-continuous-gesture}

\begin{figure}[htpb]
    \centering
    \includegraphics[width=1.0\linewidth]{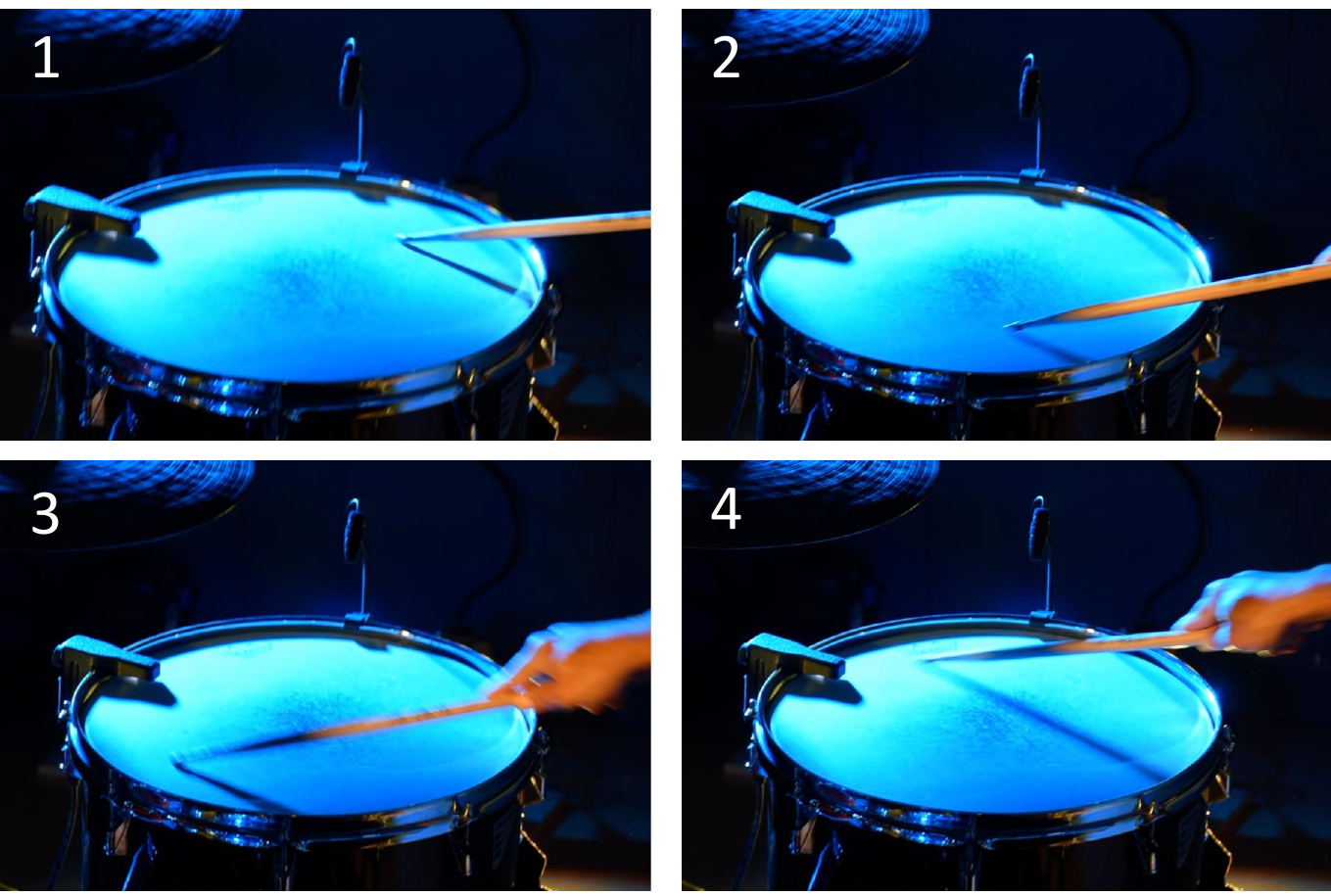}
    \caption{Drawing gesture used for continuous modulation.}
    \label{fig:drawing}
\end{figure}

The initial mapping system introduced to the musician used onset detection to trigger sounds in response to drum hits.
A representational dissonance arose when the musician pulled out a wire drum brush and asked: ``how does a brush sound?''
The resulting sound was not reflective of the gesture, unsurprisingly, as continuous brushing gestures are outside the representational scope of an onset detector.
The musician suggested that these types of continuous gestures could be mapped to synthesizer parameters, an avenue that occupied a significant portion of the study.
The continuous gesture recognition started with brushing and was expanded to include a \emph{drawing} gesture with the tip of a drumstick, visually demonstrated in Figure~\ref{fig:drawing}, and buzz stroke rolls, which the musician described as being distinguishable from controlled rolls like double strokes.

Extending the system with a neural network for continuous gesture detection expanded the representational boundaries of the onset detector, but it also introduced an artificial delineation between ``discrete'' and ``continuous'' gestures. 
This model was still ``incorrect''\textemdash the boundary is an artefact of the system, not a feature of the practice\textemdash but it was a refined model based on the musician's own practice, and one that proved useful during periods of music composition.

However, more subtle representational dissonances surfaced as we developed this approach.
For instance, even with this richer representation, the temporal envelope of the extracted feature sequences didn't always match the gestural intention of the musician.
During the final co-design session the musician suggested that a circular brushing motion, suggestive of a sinusoidal low-frequency oscillator, could be looped to provide a backing track for him to play over.
The feature envelope wasn't precise enough to support a stable loop. 
Waveshaping was introduced to shape the envelopes, but alone couldn't make up for the temporal discrepancies between gestural intention and the sonic result. 
We abandoned the looping idea (Figure~\ref{fig:toolkit-timeline}), but kept the real-time continuous feature mapping for live performance. 
Importantly, music composition was part of the design process: the gestural friction surfaced the representational limit, but it was the goal of making music that supplied a reason to do something about it.
Knowing when to push against representational limitations using technical practices and when to accept and switch into a compositional mindset emerged as an important form of judgement in our collaboration.

The productive dissonance created by the musician's practice brought representational limits into sharper relief.
Without this, the original representational framework would likely have remained as is, not because its limitations weren't known, but because there was no compelling reason to push against them from within a purely technical perspective.
It was the friction of the musician's gestural language that created the impetus to develop continuous gesture recognition.

\subsubsection{Cymbal Sensor}
\label{sec:cymbal-sensor}
A contrasting arc illustrates what happens when this balance tips too far toward technical aims. 
An IMU sensor attached to the cymbal, motivated by the performative possibilities of a visible, physical gesture, led to a spiral of interventions: each technical solution surfaced a new problem, pulling focus progressively further from musical practice. 
A visit from one of the researcher's supervisors, who had not been present for the preceding weeks of development, offered the outside perspective needed to see the spiral clearly.
We decided to scrap the sensor ahead of the recording session. 
The core idea remained compelling, but the musical rationale had receded. 
There was no musical anchor to return to because the idea had originated technically rather than emerging from the musician's practice.

\subsection{Navigating Performance-Demo Ambiguity}
\label{sec:performance-ambiguity}

A persistent ambiguity ran through the collaboration about whether the primary goal was developing technology or making music.
Despite the study's explicit motivation to support the musician's practice, the context of being a study in the first place had an effect of recentering the role of technology:

\begin{quote}
	\noindent \textbf{Researcher:} My PhD is about introducing tech to drums. [But], I'm trying to be principled and [make it] about the music.
	
	\vspace{1mm}
	\noindent \textbf{Musician:} Nah nah, it's about the tech, but we can make it also about the music. Then that means the tech has been successful.
\end{quote}

The tension between the dual objectives of the collaboration was present throughout, and was directly negotiated during a performance on the final day when a classification algorithm was producing incorrect results:

\begin{quote}
	\noindent \textbf{Musician:} I guess it's a question of whether we, or you, rather, accepting that there's a couple of errant notes here and there and sometimes it doesn't trigger. But the vibe is still there, isn't it? So on a musical level, the vibe is still working. It's just sort of more on the tech level.

	\vspace{1mm}
	\noindent \textbf{Researcher:} If it's musically there, then that's perfect.
	
	\vspace{1mm}
	\noindent \textbf{Musician:} Yeah, sweet. It's not going to be played any better, is it? Let's move on.
\end{quote}

Musical merit was identified by the musician, but attention was pulled towards a concern for demonstrating the effectiveness of the technology.
If this was indeed the only aim, then this performance would have been deemed unacceptable, but switching the frame of reference to one focused on musical merit led us to the decision to accept the performance.
This is a form of knowing-when: knowing when to evaluate using a technical lens or an aesthetic one.

However, a more subtle ambiguity arose when the researcher further questioned the musician about their experience during the performance in light of the technical issues:

\begin{quote}
	\noindent \textbf{Researcher:} I think it is interesting if you were distracted when you were playing. If you're thinking ``I'm trying to get this work and it's not really doing what I want it to.''
	
	\vspace{1mm}
	\noindent \textbf{Musician:} No, I mean, I'm sort of one of those not-perfectionist-perfectionists. I like the slight glitchiness anyway. So it's not like ``this thing's not working, and it's not doing what I wanted to.'' It's basically doing what I want it to do. And then every so often, it doesn't quite catch. And then it throws in a note when it hasn't been asked to throw in a note. But I kind of like that too. It goes in with the sort of the world we live in, a glitchy ADHD world.		
\end{quote}

The musician's response might be read as expressing a genuine aesthetic preference for the ``glitchiness'' of the mapping.
However, ``glitchy'' is an aesthetic that he hadn't mentioned during our study prior to this point---it emerged as an accommodation to the classifier failing, and was retroactively accepted as a positive.
It is difficult to assess whether this was truly a novel aesthetic discovery, or whether it was a rationalisation of a limitation that was absorbed because of the context of the study and focus on technology.

Answering this is complex.
The musician himself had framed the collaboration partly as a search for ``a new impetus, a new way to approach the next bit of writing,'' which means some degree of aesthetic influence from the technology was not just acceptable but desired.
Determining whether the aesthetic influence of the technology acts as a desirable creative catalyst or as a form of capture is central to the goals of our work. 
However, we find ourselves ill-equipped to answer this question in isolation from the musician's broader social and professional context.
We elaborate on this in the discussion.

\section{Discussion}
\label{sec:discussion}
The findings in Section~\ref{sec:productive-dissonance} surface three questions. First, what forms of tacit knowledge supported productive dissonance in our collaboration? Second, how do idiomatic representational assumptions become invisible without contact with practice---and what does productive dissonance reveal about them? Third, and hardest to answer: when technology exerts aesthetic influence on a musician's practice, how do we know whether that influence is genuinely supportive or simply going uncontested? We address each in turn.

\subsection{Knowing-When}
\label{sec:knowing-when}
\citet{ryan_knowing_2009} described a knowing-when in electroacoustic improvisation as a tacit judgement about when to intervene with an algorithm, which is similar to a type of tacit knowledge the musician described in achieving sonic specificity during live performance. 
We identify an analogous knowing-when in DMI co-design, one that supported productive dissonance in multiple ways during our study-collaboration.
Section~\ref{sec:tracing-continuous-gesture} described knowing when to push against representational limits and when to accept them and compose within their constraints.
Section~\ref{sec:performance-ambiguity} involved knowing when to evaluate using an aesthetic lens rather than a technical one.
Switching that frame was what allowed the music to take precedence over concern for demonstrating the technology's effectiveness.

A third type of knowing-when concerned the researcher's own role within the collaboration.
The researcher started the project with the intention of remaining as aesthetically neutral as possible, acting as a translator between the musician's artistic goals and the technical development.
As the study progressed, the researcher noted that taking more aesthetic initiative had a positive effect in the collaboration, even if this meant resigning from that neutrality: ``[the process is more effective] when I am also applying my musical judgements ... I'm starting a band with the musician.''\footnote{Taken from the researcher's study journal}
The musician reflected on this dynamic in kind: ``I think things sort of come naturally once you start hearing the different sound worlds.''
This process, however, raises a deeper question: did the researcher's aesthetic influence expand to fill space that the musician's broader social context might otherwise have occupied?
We return to this question below.

Knowing-when is not a design method that can be specified in advance; it is a form of judgement that develops through sustained collaborative practice.
Central to this was recognising representational assumptions that had been normalised through repeated use.

\subsection{Productive Dissonance}
The development of continuous gestures in Section~\ref{sec:tracing-continuous-gesture} sheds light on a type of representational normalisation present in our technical practice with augmented percussion instruments.
Classification algorithms, including onset detection and neural network-based detectors, are a representational foundation of the entire toolkit.
The use of these techniques as a starting point was pragmatic as much as anything else.
Onset detection solves a real problem and was used in earlier work by the researcher to facilitate mapping to MIDI-based instruments.
Similarly, classifying between head and rim hits, or continuous and discrete gestures, is a solution to a mapping problem where the output comprises discrete classes like MIDI pitch for creating melodies.
These techniques are useful, and have become somewhat idiomatic in augmented percussion and DMIs more generally. However, their usage imposes an ideological commitment on a musical practice; the act of classification enacts the concept of separable classes upon a musical practice where none may have been present before.
A designer may be well aware of the false precision of these representations, but chooses to use them anyway out of pragmatism or necessity, particularly when interfacing with tools that assert their own representational commitments (e.g., MIDI).
The risk is that through repeated use, representational convenience becomes idiomatic and eventually reified.

In our work, contact with practice, specifically a musician pulling out a wire brush and trying a mapping based on onset detection, surfaced the representational tension.
Without this productive dissonance, there was insufficient force to destabilise the representational commitments of the researcher's original design.
The musician helped push out of this.
Of course, the new continuous gesture method itself is just another classifier that introduces an artificial boundary.
Lossy representations may be unavoidable, but productive dissonance at least makes their assumptions visible---which is a precondition for designing within them consciously rather than perpetuating them.

In our study-collaboration, productive dissonance surfaced representational tensions, which led to the development of new representations that supported the musician's gestural language and composition of an album.
However, during reflection on the musical results, a more subtle question arose related to the context of the laboratory.

\subsection{The Hermetic Laboratory}
The difficulty in answering whether the glitchiness described in Section~\ref{sec:performance-ambiguity} was a positive aesthetic influence or an example of capture points to an ecological risk when working with musicians in an academic setting.
~\citet{waters_entanglements_2021} describes musical instruments as not merely ``things'' to be studied, but also a constitutive part of a larger ecosystem including performers, instruments, and environments.
With regard to musical aesthetics, ~\citet{born_social_2010} argues that treating an aesthetic as being embedded in a social context is required to accurately portray a musical practice.
Therefore, isolating a musician in a laboratory study, even with the goal of composing original music, risks severing the social dimension of a practice, potentially undermining a friction needed to distinguish genuine aesthetic development from accommodation to technological constraint.
The musician wanted the technology to provide a new creative impetus, which means some degree of aesthetic influence was desired; but without his broader practice present to evaluate that influence, would the glitchiness have been accepted?

The cymbal sensor example in Section~\ref{sec:cymbal-sensor} exemplifies a related dynamic.
A technical intervention spiral built up over the course of several sessions.
The idea had clear musical applicability, but originated from a technical possibility and became increasingly driven by a technical aim instead of a clear musical need.
Only once the supervisor visited and observed the situation were we able to break from the cycle of iterative prototyping.
In both cases the study setting worked against our aims of creating a productive dissonance with technical aims.
Technical aesthetics and institutional aims naturally filled the void left by weakened social dimensions.
This suggests that working with a musician is a necessary but insufficient condition for sustaining a productive dissonance.

This is not to say that the social dimensions were completely absent: the musician posted clips of our sessions on his social media; shared conversations he had about our work with collaborators; and we discussed public performances and a release of the album we created through traditional publishing avenues.
Additionally, the researcher himself attended musical concerts that the musician was playing at, which led to impromptu discussions related to their practice that were situated in a broader context.
All these are suggestive of a porousness of the laboratory.
Nonetheless, sustained engagement with a broader network was limited, and mostly restricted to the conclusion of the study. 
Our findings suggest deeper and sustained engagement with these dimensions is required to answer the complex, but important, questions related to technical impact on musical practice.
Situating design within an active project with real social commitments, and ones that exist outside the regular streams of academic art production, is one approach that would take this engagement further---and one that reflective practice-based methods are well-positioned to investigate further. 

\section{Conclusions}
\label{sec:conclusion}
In this work, we took up a provocation from \citet{norman_nime_2025}: ``rescue performance from the demo.'' 
This highlights a central tension in designing musical performances where technology development is a significant concern---is the performance about the music, or about demonstrating the affordances of the technology? 
When the latter occurs under the guise of composing new music, what does this reflect about the impact of a DMI on the practice it is incorporated into?

We propose that this may be indicative of technological capture, where the constraints of a DMI overwhelmingly influence a musical practice, trapping it in a prescriptive aesthetic range. 
Following recent critical work on mapping, we conducted a long-term collaboration with a professional percussionist, engaging his aesthetic as a method for creating productive dissonance with the constraints of a percussion gesture mapping toolkit---a guiding force for technical practice. 
This was partially successful, resulting in a toolkit with a novel method for mapping continuous percussive gestures using RNNs, and a ten-track album recorded with the musician, presented as an annotated portfolio.

Through a reflective practice-based approach, we surfaced a form of tacit knowledge we relate to knowing-when \citep{ryan_knowing_2009}, and identified a limitation of the academic co-design setting: isolating the social dimensions of a musician's practice risks undermining the aesthetic weight they bring to the work. 
This surfaces an irresolvable tension absent a broader social context: how do we know whether the aesthetic introduced by the technology is truly supporting the musician's aims, or whether the conditions of the study lead to its unquestioned acceptance? We put this forward as an open question, and suggest that a deep engagement with the social dimensions of a practice is a promising way to approach an answer.

\section*{Author Declarations}
The study was deemed low-risk and received ethical approval from the Queen Mary University of London EECS Devolved School Research Ethics Committee.
All AI models used during this study were trained using data recorded by the authors.
Claude Sonnet 4.6 was used to support the writing of this paper as a tool to edit text originally written by the authors.
All ideas and research code were contributed by the authors and collaborators.

\section*{Acknowledgements}
The authors would like to extend their deep gratitude to Jem Doulton for his commitment to the co-design process and for contributing his knowledge and experience.
Thank you to Carson Gant for editing the performance videos and for mixing the audio, and to Alex Bonney for mastering the tracks.
Thank you to Rodrigo Constanzo, whose DataKnot toolkit provided the foundation for this work, for the ongoing collaboration and continued conversations on drum performance and technology.

\bibliographystyle{apalike}
\bibliography{references}

\appendix
\section{Toolkit Technical Details}
\label{app:tech-details}
The technical details of the final gesture mapping toolkit that we developed are presented here.
Table \ref{tab:jem-toolkit} overviews the objects included in the toolkit.

\begin{table*}[ht]
\centering
\caption{Objects in the percussive gesture mapping toolkit.}
\label{tab:jem-toolkit}
\begin{tabular}{p{0.22\textwidth} p{0.68\textwidth}}
\toprule
\textbf{Name} & \textbf{Description} \\
\midrule
\multicolumn{2}{l}{\textit{Gesture Recognition}} \\
\midrule
\texttt{InputBrush}    & Input gesture recognition of continuous brushing gestures (uses Mel-band features). \\[2pt]
\texttt{InputBuzz}     & Input gesture recognition of buzz stroke rolls (uses onset activation signal). \\[2pt]
\texttt{InputClassify} & Input gesture recognition of discrete gestures (drum hits). \\
\midrule
\multicolumn{2}{l}{\textit{Parameter Mapping}} \\
\midrule
\texttt{MIDIReceive}     & Receives event messages from \texttt{InputClassify} and generates MIDI messages on a MIDI channel. \\[2pt]
\texttt{TimbreRemapRun}  & Receives audio features from \texttt{InputClassify} and predicts synthesizer parameters using a pre-trained neural network. \\
\midrule
\multicolumn{2}{l}{\textit{Performance}} \\
\midrule
\texttt{Looper}   & Audio looper for creating layered performances. \\[2pt]
\texttt{NoteSeq}  & Step sequencer triggered from \texttt{InputClassify}. \\[2pt]
\texttt{PresetStep}      & Steps through synthesizer channels within an instrument rack, triggered from \texttt{InputClassify}. \\
\midrule
\multicolumn{2}{l}{\textit{Utility}} \\
\midrule
\texttt{AutoGain}        & Automatically gain stages the input audio signal to a preset decibel level. \\[2pt]
\texttt{DatasetCreator}  & Records a dataset of features labelled by hit type. \\[2pt]
\texttt{TrainClassifier} & Trains a neural network classifier using a dataset recorded by \texttt{DatasetCreator}. The trained model can then be loaded into \texttt{InputClassify} for inference. \\
\bottomrule
\end{tabular}
\end{table*}

\subsection{Discrete Gesture Recognition}
\begin{figure}[htpb]
    \centering
    \includegraphics[width=1.0\linewidth]{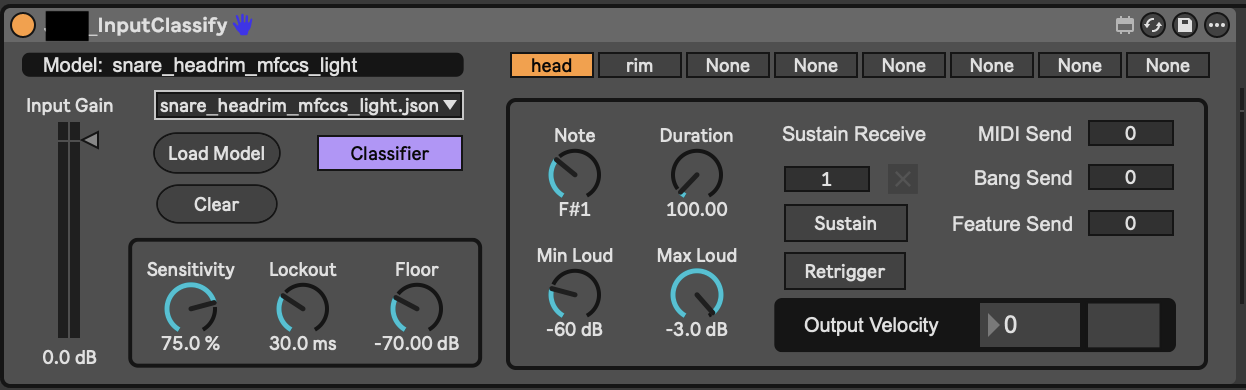}
    \caption{M4L Device for discrete gesture recognition.}
    \label{fig:toolkit-discrete}
\end{figure}

The discrete gesture recognition component of the system is event-based and utilizes onset detection to determine whether a drum has been struck at any given instance.
In our work, we primarily focused on relatively simple classification between drum head and rim\footnote{DataKnot and Senory Percussion allow for mapping of multiple different zones. This is possible in the toolkit, but is not something we explored in detail during the study.}, which we found enough in a drum-kit context to create interesting performances when mapped to different pitches, synths, and other triggers.
Figure~\ref{fig:toolkit-discrete} shows the Max4Live device, called \texttt{InputClassify}, supporting discrete gesture recognition.

\subsubsection{Method Overview}
\label{sec:discrete-method-overview}

We used the method implemented in DataKnot\footnote{\url{https://github.com/rconstanzo/data-knot}} as a starting point, which uses FluCoMa~\citep{tremblay_enabling_2021} under-the-hood.
The signal flow includes an onset detector to determine when a new even has occurred, audio feature extraction, and classification with a lightweight MLP.
The onset detector used is the FluCoMa \textit{AmpSlice} object, used throughout DataKnot, and is particularly effective for percussive signals while being lightweight and low-latency.
This method reports onsets when an activation signal computed by taking the difference between to amplitude envelope followers\textemdash one fast reacting envelope and one slow one\textemdash exceeds a specified threshold.
Parameters for onset detection are included on the UI of the Max4Live device and include a sensitivity control, which maps to both on and off threshold values, a lockout parameter which prevent re-triggering within a temporal window, and a noise floor parameter.
These are all as implemented in DataKnot.

Audio feature extraction is performed on a window of 256 samples ($\approx5.8$ms at 44.1kHz) following a detected onset.
Loudness, k-weighted, relative to full-scale (LKFS)~\citep{international_telecommunication_union_algorithms_2006} is extracted, which is normalized to a $[0,127]$ range for MIDI velocity.
Spectral features are also optionally extracted and can be sent out for timbre remapping, discussed further in Appendix~\ref{app:timbre-remapping}.

The final input features used for classification were MFCCs.
They include 13-coefficients computed on a 40-band Mel-band representation.
The 13 coefficients start from the first coefficient, dropping the 0th which encodes signal amplitude.
The minimum frequency is 200Hz and the maximum frequency is 12kHz.
MFCC frames are summarized using the mean and standard deviation.
These statistics are also computed on the first order derivative.
All statistics are weighted using loudness.
The resulting dimensionality of the feature vector is 52.

All data used for training and validating models was recorded during the co-design sessions and performed by the musician.
Datasets are recorded using the \texttt{DatasetCreator} object, which performs onset detection and feature extraction using the approach outlined above.
Classification MLPs can then be trained directly in Ableton using the \texttt{TrainClassifier} device, which utilizes \texttt{mlpclassifer\textasciitilde} from FluCoMa, or in Python\footnote{Code available on the accompanying website:\newline  \url{https://jordieshier.com/projects/aimc2026/}}. 
During the study, Python was primarily used for training, which was useful for iterating model design and evaluation, which provided us with additional options for data pre-processing.

Prior to training, all features are normalized.
Class imbalance is addressed in Python using the synthetic minority over-sampling technique (SMOTE)~\citep{chawla_smote_2002}, which balances classes by generating new data points for minority classes.
For each data point, a random neighbour is selected and a new sample is generated in the direction of that neighbour by multiplying the difference between their feature vectors by a random number between 0 and 1, and then adding the result to the original sample.

An MLP classifier is then trained using the same approach used in FluCoMa, replicated in Python.
This approach treats classification as a regression problem and optimizes the mean squared error (MSE).
Output activations pass through a sigmoid non-linearity and loss computed on a one-hot encoded ground truth vector.
The network is optimized using stochastic gradient descent.

In our study, we focused on simple classifications between the head and rim, although the Max4Live device can support multiclass classification of up to eight classes.
A lightweight MLP with a single hidden layer of two neurons and a ReLU activation was sufficient for this prediction.
Models were trained for a maximum of 1000 epochs with a learning rate of \num{1e-3} and a momentum of 0.9.
Early stopping is applied when loss on withheld 20\% validation split doesn't decrease for 20 epochs. 

FluCoMa's \texttt{mlpclassifer\textasciitilde} is used for inference in the Max4Live device, and each class can output one of three different messages, which are routed internally within the toolkit: 1) trigger only; 2) MIDI note and velocity; 3) spectral features.

\subsection{Continuous Gesture Recognition}
\label{app:continuous-mapping}
\begin{figure}[htpb]
    \centering
    \includegraphics[width=1.0\linewidth]{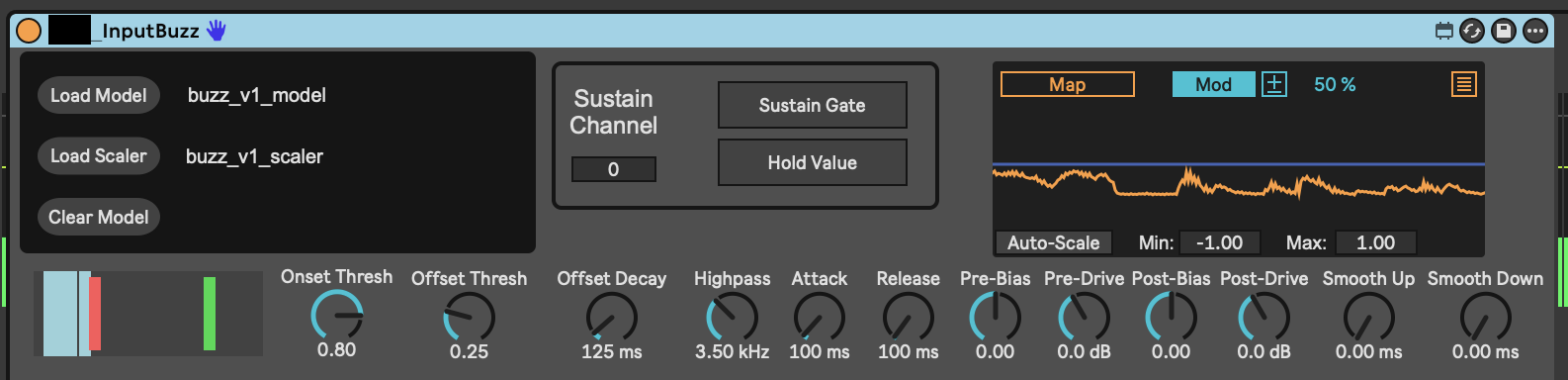}
    \caption{M4L Device for continuous gesture recognition. This device is \texttt{InputBuzz}; however, \texttt{InputDrush} shares the same interface design.}
    \label{fig:toolkit-continuous}
\end{figure}

We focus on two different types of continuous gestures for detection: 1) gestures resulting in a continuously sustained sounds like brushing or ``drawing'' on a drum membrane with the tip of a stick;  and 2) buzz stroke rolls or drags, which are distinguished from more consistent strokes like single or double strokes.
There are two different devices for each type of continuous gesture---\texttt{InputBrush} and \texttt{InputBuzz}---they differ only in their input audio representation, described below.
In addition to gesture recognition, a continuous loudness feature is extracted, optionally gated by the continuous gesture prediction, and can be mapped within Ableton.
Figure~\ref{fig:toolkit-continuous} shows the Max4Live device for continuous gesture recognition.

\subsubsection{Input Representations}
\label{app:cont-features}

Brush and drawing gesture recognition use a Mel-spectrogram representation computed using an FFT with a size of 512 and a hop-size of 256 samples at a sampling rate of 48kHz.
Each frame is windowed with a Hann window prior to the FFT.
Mel-band magnitudes are then computed by multiplying the magnitude spectrum with triangular filters.
We use 32 Mel-spaced filters with a minimum frequency of 1kHz and a maximum frequency of 20kHz.
The continuous gestures are characterized by continuous noise, which we aim to capture with this representation.
The resulting vector is then normalized by magnitude to preserve relative energy distribution across Mel-bands.
Mel-band magnitudes are computed from the FFT results as follow:

\begin{align}
    X^{\prime}_k &= \frac{4}{N}|X_k|\\
    E &= \frac{1}{2}\sum_k{X^{\prime}_k}\\
    \hat{m}_j &= m_j\frac{E}{\sum_j{m_j}}
\end{align}

\noindent where $m_j = \sum_k{X^\prime_k f_{jk}}$ and $f_{jk}$ are 
the Mel filterbank weights for the $j$th Mel-band and $k$th FFT bin.
The final Mel-bands are converted to a decibel scale.

Since the buzz stroke rolls are better characterized by temporal changes, we use an onset activation representation for these gestures.
This activation signal is computed using the same method as the FluCoMa \emph{AmpFeature} algorithm.
First the input signal is high-pass filtered at 2kHz.
This signal is then rectified, converted to a decibel scale, and floored:

\begin{equation}
    v[n] = \max(-70, 20\log_{10}({|x[n||}))
\end{equation}

\noindent This amplitude signal is then passed through two parallel envelope followers, a fast envelope follower and a slow one:

\begin{equation}
    y[n] = y[n-1] + \alpha \left(v[n] - y[n-1]\right)
\end{equation}
\noindent where
\begin{equation}
    \alpha = \begin{cases} \alpha_{\text{up}} & \text{if } v[n] > y[n-1] \\ \alpha_{\text{down}} & \text{otherwise} \end{cases}
\end{equation}

\noindent where the fast envelope follower has $\alpha_{\text{up}}=1/3$ and $\alpha_{\text{down}}=1/300$, and the slow envelope follower has $\alpha_{\text{up}}=\alpha_{\text{down}}=1/2205$.
These values are selected from the defaults in DataKnot.
A difference envelope is then computed by subtracting the slow envelope from the fast envelope.
These three envelopes are downsampled by a factor 64 and concatenated into a 3-dimensional time series.

Both of these representations are normalized using values computed on a training dataset.

\subsubsection{Neural Network Architecture and Training}
The recognition network utilizes a gated recurrent unit (GRU)~\citep{cho_properties_2014} recurrent neural network with an architecture similar to the one used by~\citet{engel_ddsp_2020}.
Our version has a single GRU layer followed by an MLP.
For our study, we used a GRU with a single hidden layer of size 16, and an MLP with three hidden layers, each with a size of 16 neurons.
The MLP use ReLU activations on the hidden layers and the final 1D output has a sigmoid activation.

Training data was collected during the co-design sessions, and regions of continuous gestures manually labelled using Sonic Visualiser~\citep{cannam_sonic_2010}.
Models are trained in Python using binary cross entropy loss for 250 epochs with a batch size of 16 using an Adam optimizer with a learning rate of \num{1e-3}.

\begin{figure}[t]
    \centering
    \begin{subfigure}{\linewidth}
        \centering
        \includegraphics[width=0.99\linewidth]{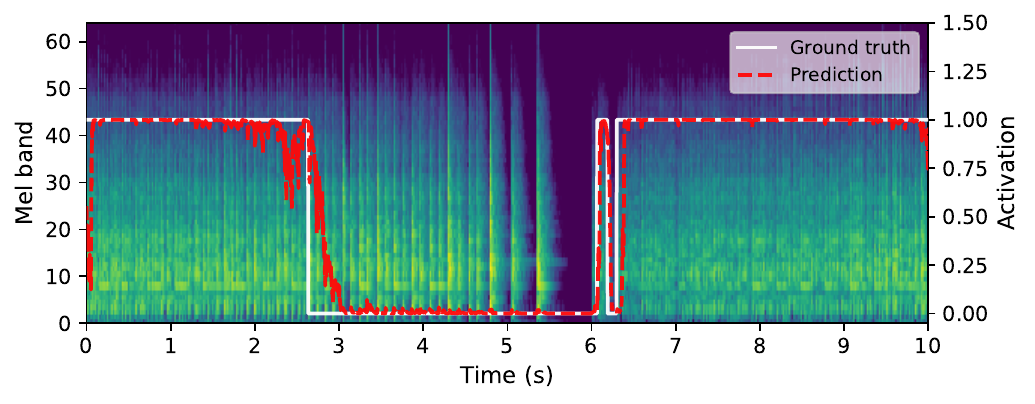}
        \caption{Continuous gesture example for a buzz stroke roll on a snare drum transitioning to a double stroke roll, then a buzz stroke roll starting again.}
        \label{fig:buzz-rnn}
    \end{subfigure}

    \vspace{1em}

    \begin{subfigure}{\linewidth}
        \centering
        \includegraphics[width=0.99\linewidth]{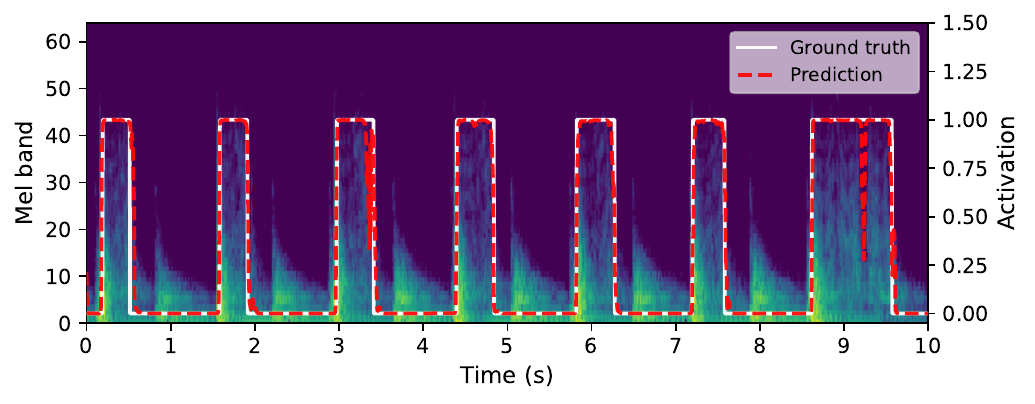}
        \caption{Continuous gesture example for drawing gestures interleaved with hitting on a floor tom. The model ignores the hitting gestures and responds to the drawing.}
        \label{fig:brush-rnn}
    \end{subfigure}

    \caption{Example segments from continuous gesture prediction. Each plot shows a spectrogram of the audio input, the ground truth labels, and the model prediction.}
    \label{fig:continuous-figures}
\end{figure}

Figure~\ref{fig:continuous-figures} shows example outputs of two different models, one trained for buzz stroke roll recognition and the other for brush/drawing recognition.

\subsubsection{Real-time Implementation and Mapping}
We used the RTNeural Plugin\footnote{\url{https://github.com/spluta/RTNeural_Plugin}} for Max/MSP, which builds on the RTNeural neural network inferencing library developed by~\citet{chowdhury_rtneural_2021}.
Input features are extracted in Max/MSP using the same methods described above in Appendix~\ref{app:cont-features} at a frame rate of 750Hz (equivalent to a hop size of 64 samples).
Neural network predictions occur at this frame rate.
Optionally, the output stream of predictions is smoothed with an envelope follower with a parameterized exponential decay rate.
Onset and offset thresholds are also parameterized on the UI of the Max4Live object.
The object outputs a message similar to a sustain pedal MIDI CC value, which is high when in a continuous gesture and low otherwise.

Loudness is extracted continuously on the input signal, scaled to a $[-1,1]$ range, and can be mapped to any parameter that can be modulated in Ableton.
A highpass filter is applied before loudness computation with a parameterized cutoff frequency.
Shaping is applied to the loudness feature:

\begin{equation}
    y[n] = \text{clip}\!\left(\left(\alpha_1 + \tanh\!\left((\alpha_0 + x[n]) \cdot \beta_0\right)\right) \cdot \beta_1\right)
\end{equation}

\noindent where $\text{clip}(x) = \max(-1, \min(1, x))$, $\alpha_0$ and $\alpha_1$ are pre- and post-DC offsets, and $\beta_0$ and $\beta_1$ are pre- and post-gain factors, and $x[n]$ is the scaled loudness signal.
To handle different ranges of loudness signals, we added an optional adaptive scaling that samples the loudness every 50ms and computes minimum and maximum values for scaling on the previous 20 values.
The resulting signal is then smoothed using an exponential envelope follower with controllable attack and release times.

Finally, there are options for how the feature behaves depending on whether a continuous gesture is detected.
By default, the feature is always sent, regardless of the detected state.
A \emph{sustain gate} can be applied which allows active modulation only during a continuous gesture, and when a gesture stops the modulation can be set to drop to zero or held at the previous value.
Attack and release times control how the signal behaves during transitions between continuous and non-continuous detected states.

\subsection{Timbre Remapping}
\label{app:timbre-remapping}

\begin{figure}[htbp]
    \centering
    \includegraphics[width=1.0\linewidth]{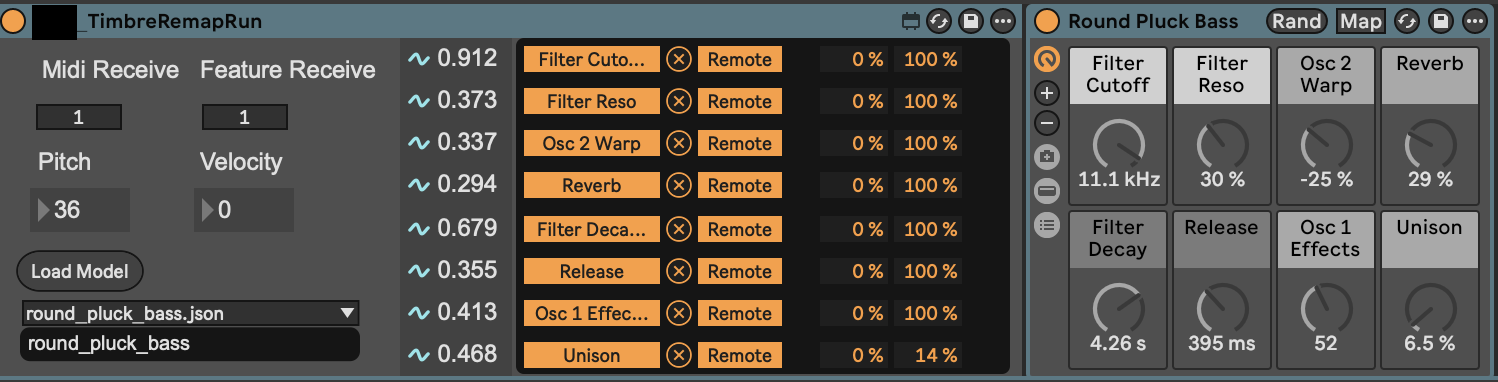}
    \caption{M4L Device for Timbre Remapping Inference}
    \label{fig:toolkit-timbreremap}
\end{figure}

The timbre remapping component of the toolkit is based on the method described in~\citep{shier_designing_2025}, which we refer the reader to for full technical details. 
Briefly, an offline process generates a corpus of synthesizer presets paired with their associated audio features, which is used as training data for a neural network. 
At runtime, the network predicts synthesizer parameters from spectral audio features extracted at detected onsets. 
For this study, we adapted this approach for integration with Ableton Live as the \texttt{TimbreRemapRun} Max4Live device (see Table~\ref{tab:jem-toolkit}). 
Figure~\ref{fig:toolkit-timbreremap} shows the Max4Live device for timbre remapping inference
This object recieves MIDI and spectral features from an \texttt{InputClassify} device, neural network inference is performed with the spectral features, producing synthesizer parameters, which are mapped to parameters in Ableton\footnote{There are eight mappable output parameters, designed to work with macros on an instrument rack.}, then the synth is triggered with the input MIDI event.

\subsection{Performance Devices}
The toolkit includes three performance devices to support musical structure within a live performance context.

\begin{figure}[htpb]
    \centering
    \includegraphics[width=1.0\linewidth]{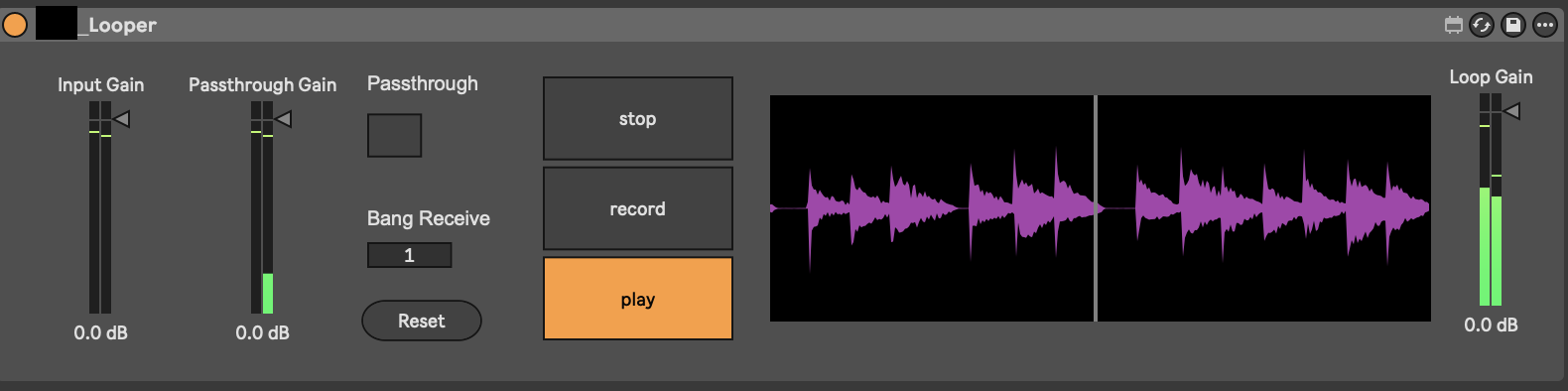}
    \caption{M4L Looper Device}
    \label{fig:toolkit-looper}
\end{figure}

The \texttt{Looper} device wraps the \texttt{karma\textasciitilde} looper\footnote{\url{https://rodrigoconstanzo.com/karma/}}, and performs loop actions in response to a trigger from \texttt{InputClassify}.
The looper here is a simple state machine that cycles from standby, to recording, to playback, and then back to standby.
It was used in one track and enabled the musician to loop the output of a synthesizer and then play drums overtop.
Figure~\ref{fig:toolkit-looper} shows he Max4Live looper device.

\texttt{NoteSeq} is a performance device for creating a sequence of pitch values.
The object is triggered via MIDI received from \texttt{InputClassify}.
Optionally, the sequence can be stepped forward only when a specific MIDI note has arrived, enabling the musician to sit on a single note in the sequence until a gesture is received to change the loop.

\begin{figure}[htpb]
    \centering
    \includegraphics[width=1.0\linewidth]{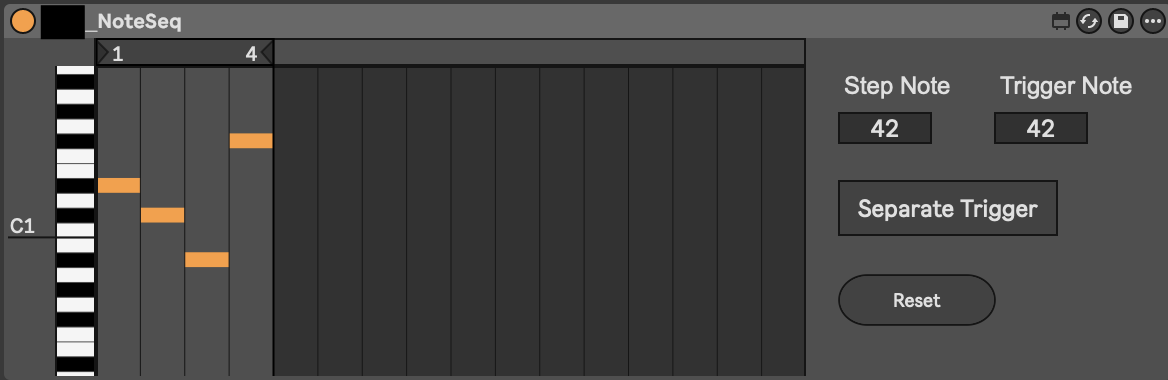}
    \caption{M4L Note Sequencing Device}
    \label{fig:toolkit-noteseq}
\end{figure}

\begin{figure}[h!]
    \centering
    \includegraphics[width=0.5\linewidth]{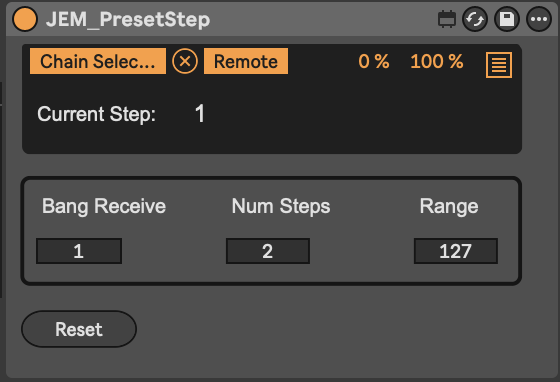}
    \caption{M4L Preset Step Device}
    \label{fig:toolkit-presetstep}
\end{figure}

The final device is \texttt{PresetStep}, which was designed to allow for larger sonic changes like switching the synthesizer preset.
This device operates as a state machine that steps when a trigger is received from an \texttt{InputClassify} device.
The output state is configurable to have between 2 and 127 states, and is mappable within Ableton.
Used in conjunction with an instrument rack with multiple chains, mapping the chain selector control enabled us to switch between different instrument chains from a drum trigger, which was useful for creating different musical sections within a piece.
Figure~\ref{fig:toolkit-presetstep} shows the device.

\end{document}